## HTEM data improve 3D modelling of aquifers in Paris Basin, France

G. Martelet[1], P.A. Reninger[1], J. Perrin[1], B. Tourlière[1]

[1] Bureau de Recherche Géologique et Minière (BRGM), UMR 7327, 3 avenue Claude Guillemin, BP 36009, 45060 Orléans Cedex 2

In Paris Basin, we evaluate how HTEM data complement the usual borehole, geological and deep seismic data used for modelling aquifer geometries. With these traditional data, depths between ca. 50 to 300m are often relatively ill-constrained, as most boreholes lie within the first tens of meters of the underground and petroleum seismic is blind shallower than ca. 300m. We have fully reprocessed and re-inverted 540km of flight lines of a SkyTEM survey of 2009, acquired on a 40x12km zone with 400m line spacing.
The resistivity model is first "calibrated" with respect to ca. 50 boreholes available on the study area. Overall, the correlation between EM resistivity models and the hydrogeological horizons clearly shows that the geological units in which the aquifers are developed almost systematically correspond to relative increase of resistivity, whatever the "background" resistivity environment and the lithology of the aquifer. In 3D Geomodeller software, this allows interpreting 11 aquifer/aquitar layers along the flight lines and then jointly interpolating them in 3D along with the borehole data.
The resulting model displays 3D aquifer geometries consistent with the SIGES "reference" regional hydrogeological model and improves it in between the boreholes and on the 50-300m depth range.

**Introduction**

In France, in basin environment, regional geometries of aquifers are traditionally derived from the interpolation of borehole data combined with the intersection of the geology with the topography. For deeper parts of the aquifers, seismic data, generally acquired for O&G purposes, are also often used in order to constrain the geological geometries at depths greater than 300 m, as these data are "blind" at shallower depths.
With these traditional approaches aquifer geometries at depths between ca. 50 to 300 m are often relatively ill-constrained as most boreholes lie within the first tens of meters of the underground.

In the meantime, HTEM methods have rapidly evolved in the last decades and nowadays enable fast mapping of the resistivity over extensive areas, with investigation depths usually ranging from around 150 m to 300-500 m, and resolutions addressing regional geology imagery (e.g. Legault, 2015). Because the resistivity is sensitive to both the nature of the rocks, their porosity, their clay content and the presence and mineralization of the water, EM imagery is suitable for providing information about aquifer structures (and water quality) (e.g. Siemon et al., 2009).

In 2009, as part of a regional research project, BRGM (the French geological survey) conducted a HTEM survey in the south-center of Paris Basin, west of the city of Vierzon, on a 40 x12 km area (Figure 1). SkyTEM ApS. surveyed 540 km of flight lines, at a 400 m line spacing. The surveyed area encompasses geological formations ranging from the Upper Jurassic (in the south) to the Miocene (in the north), all of them gently dipping towards the north (Figure 1). In this sedimentary pile hydrogeologists inventory 5 major aquifers, which were already modelled at the scale of the administrative district (see cross-section in Figure 1) within the SIGES Centre database (Salquebre, 2012). This database is the reference for decision making in the domain of water supply at the regional level.

Present study investigates the potential of HTEM data to complement the usual borehole data and geological maps used for modelling aquifer geometries: how do EM soundings react in such sedimentary environment? Do the aquifer/aquitard mark themselves? Can the resistivity model constrain the hydrogeological geometries and/or improve the existing SIGES model?

**Material and method**

From north to south (and from the surface to deeper levels), 5 main geological ensembles outcrop in the study area: at the top, thin Quaternary alluvial deposits mainly represented in the flat northern plain cover the Cenozoic "Sologne sands and clays" which are imbricated and rapidly-changing alternations of small and ill-known aquifers / aquitards. Below, the first well identified few-meters-thick aquifer is called "Beauce" composed of limestone to more marly facies. It is separated from the underlying thick Cretaceous chalk aquifer by strongly weathered chalk which forms a blanket of clay with flints. Below, Ceno-Albian marls and clays overlay a rather thin Cenomanian sand aquifer. Next underlying aquifer is well developed, composed of Albian sands. It overlays Aptian to Tithonian marly and clayey facies which isolate the deeper aquifer considered in the study: the Upper Jurassic massive limestones.
This 11 aquifer / aquitard alternation constitutes a simplified geological pile which is the basis of SIGES regional hydrogeological model. We re-use this framework in our study: ca. 50 boreholes from which only 7 exceed the depth of 50 m are inventoried in the study area; they are coded according to the 11-layer pile.

Concerning the HTEM data, the spacing between each EM sounding along the flight lines is approximately 50 m and the nominal height of the loop was about 50 m above the ground. For this survey, the low moment (LM) had a magnetic moment of approximately 3760 Am² with time gates from 11 µs to 115 µs and the high moment (HM) had a magnetic moment reaching 140000 Am² with time gates from 73 µs to 9 ms. This configuration allows a penetration of ca. 200 m (depending on the nature of the underground) and a resolution little better than 5 m at the surface.

Since 2009, in order to achieve the best possible quality of data, BRGM has thoroughly reprocessed this survey (partial resampling of decays, denoising of data using the SVD approach (Reninger et al., 2011), use of differential GPS rather than laser/radar to determine a precise ground clearance, correction of the effective tilt of the frame (Reninger et al., 2015).

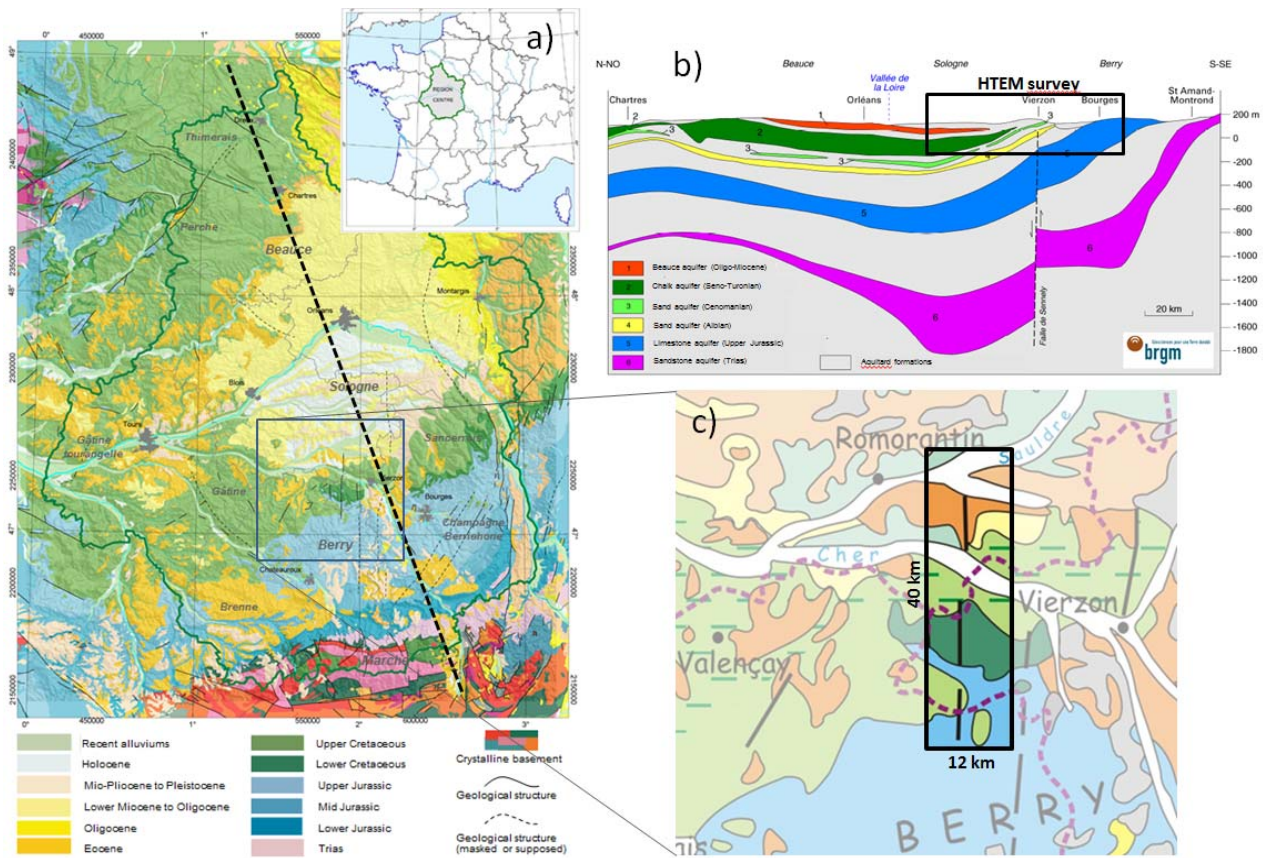

*Figure 1*- a) Location and regional geology of the study area; b) hydrogeological transect extracted from SIGES regional hydrogeological model (after Salquebre, 2012) (located by a dashed line on map (a)); c) close-up on the geology of Vierzon area, where the 40 x 12 km HTEM survey was conducted - its footprint is also approximately located by a black box on section (b).

The EM dataset was inverted using the Spatially Constrained Inversion algorithm (SCI) (Viezzoli et al., 2008). In this algorithm a pseudo-3D inversion is performed; vertical and lateral (along and across the flight lines) constraints are applied on 1D earth models divided into *n* layers, each being defined by a thickness and a resistivity. Results were first obtained with a smooth inversion (25 layers) and a test was made with a 9 layer inversion. Few layers better discriminate geological interfaces and resistivity contrasts but at the expense of visual continuity of layers which is generally better in smooth inversion.

After a preliminary geophysical evaluation / optimization of the inversion results, we interpolated (by inverse distance) the resistivity and thicknesses of each inverted layer separately, in order to preserve as much vertical resolution as possible. Both the smooth and few layer resistivity models, either in 1D or 3D were imported in 3D Geomodeller software, together with the 1:50.000 geological map draped on a 50 m DTM, and the ca. 50 boreholes available on the HTEM survey area.

**Results**

In a first stage of interpretation, in order to characterize the resistivity response of the targeted aquifers / aquitards, a detailed comparison between the borehole logs and the 1D resistivity soundings was achieved. An example of such local benchmarking of EM responses versus geology and hydrogeology is displayed in Figure 2.

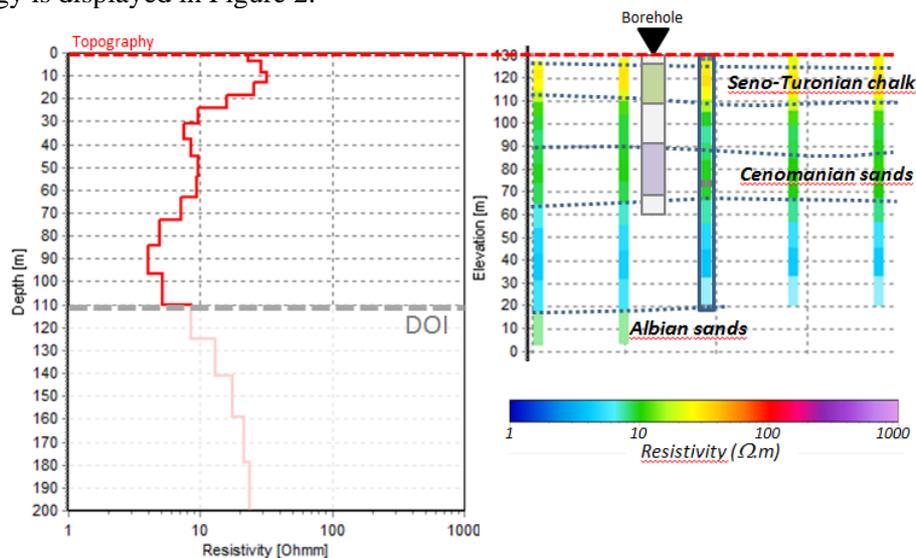

*Figure 2- A borehole is taken as reference to "calibrate" the interpretation of the EM resistivity responses along the flight lines.*

Overall, the correlation between EM resistivity models and the geology recognized in boreholes was good. It showed that the geological units in which the aquifers are developed almost systematically correspond to a noticeable relative increase of resistivity. This is not surprising as both the effect of water (when weakly mineralized), the effect of porosity (expected in aquifers) and less clay, tend to increase the resistivity. However we could experience it very clearly, whatever the lithology of the aquifer and whatever the range of resistivity of the terrains surrounding the aquifer. This is illustrated in Figure 3 where we observe that the "background resistivity" between the four concerned aquifers varies over at least two decades of resistivity. And whatever the background resistivity range, the aquifers show up as more resistive layers than their environment. In particular, the Cenomanian sands have a very peculiar behaviour: they hardly reach 10 $\Omega$.m but this is still significantly higher than the surrounding clays and marls which have resistivity of a few ohm.meters.

In comparison to the SIGES reference hydrogeological model, the EM resistivity was generally in good accordance with the pre-existing interfaces (left part of Figure 3), but for depths greater than ca.100 m and in places where few boreholes were available (right part of Figure 3), the resistivity proved very appropriate to improve the aquifer / aquitard geometries

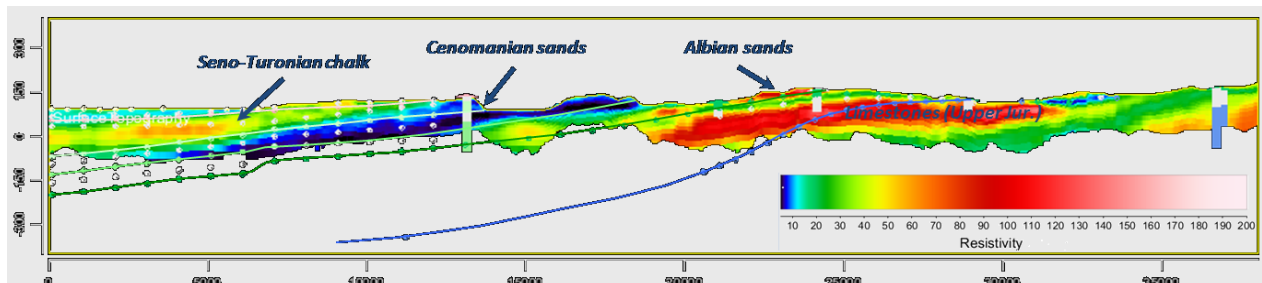

*Figure 3- Colored interfaces represent aquifer/aquitard geometries derived from the pre-existing SIGES hydrogeological model. For deeper parts of the model and in places where few boreholes were available (right part of the section), the resistivity allowed improving the hydrogeological interfaces.*

Two aquifers however exhibited ill-defined resistivity signatures: i) as expected, the "Sologne sands and clays" display resistivity values ranging from ca. 10 to 40 $\Omega$.m split into more or less lenses of resistivity: we expect that this corresponds to small sandy aquifers within a clayey background (this has been reported locally in the field, but never characterized on a regional scale); ii) from place to place the "Beauce" aquifer exhibits a weak increase of resistivity, and sometimes even looks more conductive than its surrounding: this is likely related to its variations between true limestone to more marly facies.

Once the ranges of resistivity of the 11 geological / hydrogeological layers had been characterized, the HTEM resistivity model has been thoroughly interpreted in profile along the flight lines and then jointly interpolated in 3D along with the borehole data. Resulting geometries are shown in Figure 4.

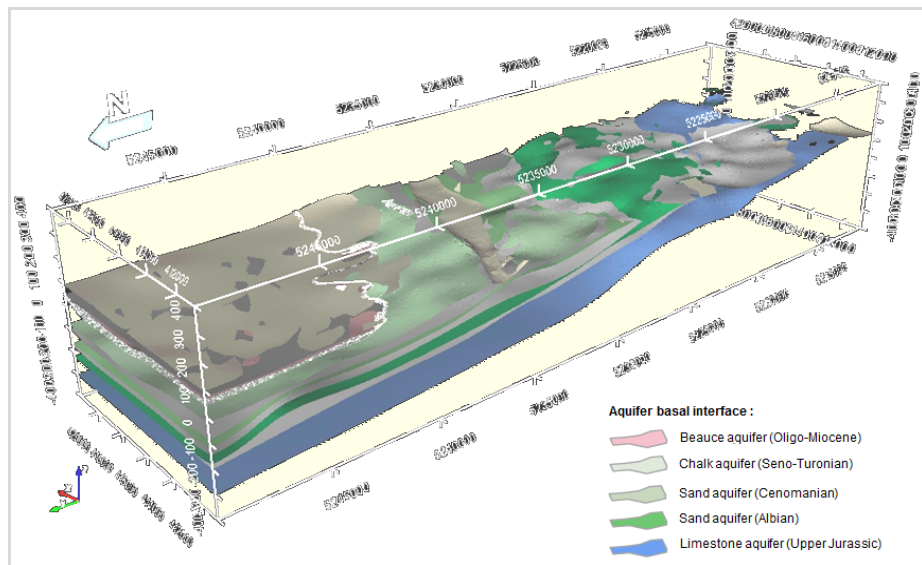

*Figure 4*- *3D geometries of aquifers derived from the interpretation of ca. 50 boreholes and HTEM data, interpolated in 3D Geomodeler software. Model extent: 40x12km - 300m thick (vert. exag. x10).*

**Conclusions**

Starting from the SIGES existing reference regional hydrogeological model, this work allowed i) to evaluate how HTEM resistivities react in a sedimentary environment, ii) to qualify the way how aquifer/aquitard mark themselves and iii) in accordance with boreholes, allowed to derive consistent 3D aquifer geometries which improve the SIGES model, particularly on the 50-300 m depths.

**Acknowledgements**

HTEM data used in this study were jointly funded by UE FEDER, Région Centre and BRGM.